# Lasing from complete set of topological states in two-dimensional photonic crystal structure

Changhyun Han[1,2], Minsu Kang[1,2], and Heonsu Jeon[1,2,3,*]

[1]Department of Physics and Astronomy, Seoul National University, Seoul 08826, Republic of Korea

[2]Inter-university Semiconductor Research Centre, Seoul National University, Seoul 08826, Republic of Korea

[3]Institute of Applied Physics, Seoul National University, Seoul 08826, Republic of Korea

[*]email: hsjeon@snu.ac.kr

**Abstract**

Recently, topologically engineered photonic structures have garnered significant attention as their eigenstates may offer a new insight on photon manipulation and an unconventional route for nanophotonic devices with unprecedented functionalities and robustness. Herein, we present lasing actions at all hierarchical eigenstates that can exist in a topologically designed single two-dimensional (2D) photonic crystal (PhC) platform: 2D bulk, one-dimensional edge, and zero-dimensional corner states. In particular, multiple topological eigenstates are generated in a hierarchical manner with no bulk multipole moment. The unit cell of the topological PhC structure is a tetramer composed of four identical air holes perforated into an InGaAsP multiple-quantum-well epilayer slab. A square area of a topologically nontrivial PhC structure is surrounded by a topologically trivial counterpart, resulting in multidimensional eigenstates of one bulk, four side edges, and four corners within and at the boundaries. Spatially resolved optical excitation spontaneously results in lasing actions at all nine hierarchical topological states. Our experimental findings may provide insight into the development of sophisticated next-generation nanophotonic devices and robust integration platforms.

Inspired by *topological insulators* (TIs) for electrons, topological concepts and ideas have been rapidly transferred to other classical waves, such as electromagnetic[1], acoustic[2], and vibrational waves[3], because the topological nature offers many sophisticated and exotic properties. Among others, the most distinct and fascinating topological features are *topological edge states* (TESs), i.e. physical eigenstates that emerge at topological edges or boundaries. The existence of TESs is guaranteed by *bulk-boundary correspondence*[4], which states that a $D$-dimensional insulating TI bulk possesses ($D–1$)-dimensional conducting TESs. A topological symmetry generates and classifies TESs as well as protects their characteristics, thereby offering robust properties.

In *topological photonics,* which addresses topological effects on electromagnetic waves, two-dimensional (2D) structures have been used as the main platforms because they not only facilitate structural realisation, but also enable studies of photonic analogues of many fascinating 2D quantum electronic phenomena, such as quantum (spin) Hall effects[5]. In fact, most topological photonics studies thus far have addressed one-dimensional (1D) TESs existing in a topological 2D platform with the main emphasis on their propagation (or conducting) characteristics[6-14]. As a confined topological state, zero-dimensional (0D) TESs in a topological 1D structure have been studied, which have expanded our view point on TESs from wave transportation to confinement and trapping[15-20]. The most studied topological 1D structure is based on the Su–Schrieffer–Heeger (SSH) model[21]; however, the model is highly simple and cannot provide sufficient structural variations required for various fundamental studies and sophisticated application devices.

Recently, higher-order (or lower-dimensional) TESs have been discovered, suggesting that a $D$-dimensional TI bulk can possess TESs of dimensions less than $D–1$[22-24]. The possibility of lower-dimensional topological phases based on bulk multipole moments has been predicted[22] and was followed by experimental demonstrations in various systems, such as mechanical resonators[25], electrical circuits[26,27], and photonic microring arrays[28]. An alternative to form higher-order insulating TESs without bulk multipole polarisations has been suggested[24], where a nontrivial topology induced by bulk dipole moments generates higher-order TESs in a hierarchical manner. Specifically, a 2D bulk with nonzero polarisations induces 1D edge states, which consequently generates a 0D corner state where two 1D edge states with nonzero polarisations encounter[29-34]. This type of higher-order TESs is useful when multipole polarisations are absent or difficult to realise.

In photonics, extreme light confinement using photonic crystals (PhCs) or surface plasmons is a fascinating research area that may result in strongly enhanced light-matter interactions in an ultimately small scale, which will significantly affect the development of high-density photonic integrated circuits. In this context, the recent experimental demonstration of topological 0D corner states in a 2D PhC slab structure[34] is noteworthy, although stimulated emission at the topological 0D state was not demonstrated—even at a cryogenic temperature. It is noteworthy that all TES-based laser devices demonstrated hitherto[18,19,35-38] rely on the lowest-order ($D$–1)-dimensional TESs.

In this study, we demonstrate lasing actions at all hierarchical TESs (bulk; edge; corner states) that exist in a topological 2D PhC structure. The coexistence of multiple hierarchical TESs in a topological PhC structure has been suggested theoretically[31] and proven experimentally in the microwave range[32,33]. We demonstrate lasing at multiple topological states in the optical frequency range by realising such a 2D PhC topological structure in an InGaAsP-based multiple-quantum-well epilayer using nanofabrication technology. The resultant lasing modes are not only small, but also multiplexed both spectrally and spatially, which are highly desirable and useful for applications such as high-density photonic integrated circuits or sophisticated photonic sensors.

## Results

**Topological 2D PhC structure**

In designing a 2D topological structure, it is natural to generalise the celebrated 1D SSH model[30,39]. We implemented our 2D SSH structure in a PhC slab platform, in which photonic atoms were tetramerised to form a unit cell, as shown in Fig. 1(a); it is noteworthy that our structure exhibits a hole-in-slab configuration, not a cylinder-in-air configuration[32,33]. Square-shaped holes (side length: $s = 0.33a$) were drilled into a dielectric slab (thickness $t = 230$ nm; refractive index $n = 3.4$) to form a unit cell (lattice constant: $a = 500$ nm) composed of four air holes arranged in a square symmetry. The four holes were spaced equally apart by distance $d$ from the centre of the unit cell in the $x$- and $y$-directions; $d$ is the key parameter that controls topological details of the structure. We assume that the modes are transverse-electric (TE) as the slab waveguide supports the fundamental TE-guided mode only.

Figure 1(b) shows the photonic band structures calculated for three unit cells with different $d$ values: $d_1 = 0.25a - \Delta$, $d_2 = 0.25a$, and $d_3 = 0.25a + \Delta$, where $\Delta$ was selected to be $0.07a$. The calculations were performed using a full three-dimensional finite-difference time-domain (FDTD) method (see Methods). When $d = d_2$, we obtained a gapless double-degenerate band structure, which was due to band folding as the lattice constant ($a$) was twice the actual periodicity ($2d$). For $d = d_1$ or $d_3$, the bandgap opens. In fact, the two band structures are identical because $d_1$ and $d_3$ are complementary ($d_1 + d_3 = a/2$); two different unit cells can be selected to produce the same PhC structure—see the insets in Fig. 1(b). Nonetheless, they possess different bulk polarisations. It is clear that the bulk polarisation is nonzero (topological) for $d > d_2$, which results in a TES at the boundary of a finite structure; however, it becomes zero (trivial) for $d < d_2$ (see Supplementary Information for details).

**Hierarchical topological states**

Typically, the TESs of the 2D SSH model structures are calculated by the tight-binding method[28,39]. However, the latter is not suitable for PhC-based SSH structures because photonic interactions easily extend beyond the nearest-neighbour atoms; the only exception may be when an array of PhC cavities is employed[20]. Therefore, we conducted FDTD simulations to accurately simulate the TESs in our structure. As shown in Fig. 2(a), we first investigated 1D edge states that may manifest themselves at the interface where two topologically distinguished PhC structures ($d_1$ and $d_3$) are joined together. The region surrounded by the red solid lines in Fig. 2(a) represents the unit cell for the FDTD calculation with the Bloch boundary condition in the *x*-direction. The resultant projected band structure is shown in Fig. 2(b), which reveals a new single band inside the bulk band-gap—see the solid blue line in the figure. To identify the nature of this band, we conducted a mode profile calculation at an arbitrary point indicated by the star on the band ($k = 0.35 \times 2\pi/a$). As shown in Fig. 2(c), its modal distribution within the unit cell is characterised by a strong localisation at the interface, confirming that it is a 1D TES propagating along the topological line interface along the *x*-direction.

After confirming the existence of the 1D edge state, which is the lowest-order TES of the topological 2D PhC structure, we performed another numerical simulation to verify the coexistence of the higher-order 0D TES at the corner. The simulation structure was designed such that the topologically nontrivial PhC structure ($d_3$) was surrounded by the topologically trivial structure ($d_1$). Figure 2(d) schematically shows a 90° corner interface of the structure. Despite the lack of quantised bulk quadrupole moment, which is apparent from the fact that only a single band exists below the bulk bandgap, our structure may still possess a corner charge $Q$ induced by two topologically nontrivial quantised dipole moments at the side edges, $p_x$ and $p_y$, in a hierarchical manner[29-34]. FDTD simulations reveal a distinct resonant mode outside the edge state band but still inside the bulk bandgap—see the dotted red line in Fig. 2(b). It is noteworthy that the 1D edge state band is gapped, which distinguishes our PhC structure from conventional topological insulators, such as Chern, spin Hall, and valley Hall insulators, where the first-order TESs are gapless. In our opinion, this gap in the first-order TES enables the simultaneous existence of higher-order TESs. The field distribution simulated for the state, which is shown in Fig. 2(e), clearly demonstrates that the mode is strongly localised at the corner, a direct indication that it is indeed a confined 0D corner state.

**Lasing from three topological states**

To experimentally identify these multidimensional topological photonic states and explore their application possibility as photonic devices, we fabricated the proposed structures using an InGaAsP multiple-quantum-well (MQW) epilayer (see Methods for detailed fabrication procedures). Although a large $\Delta$ value is preferred for a wide bandgap and robust TESs, extremely narrow (in width) ribbons are required between the square-shaped air-holes, rendering device fabrication formidable, if not impossible. As a quick solution for the structural dilemma, we investigated the extreme condition where $d = s/2$ (or $\Delta = 0.085a$ for a fixed $s = 0.33a$) such that the four square holes in the unit cell abutted to each other, as shown in Fig. 2(f). It is noteworthy that the topological nature and associated topological properties discussed thus far are not altered even in this extreme situation because $\Delta > 0$, *i.e.* the selected $\Delta$ value does not cross the topological phase transition point. Figure 2(g) shows a scanning electron microscopy (SEM) image of the fabricated device. The inset in the figure shows an amplified image of a corner boundary region.

The fabricated device structure was optically excited with a 1064 nm laser diode in pulsed mode (at 500 kHz frequency for 10 ns). The excitation laser beam was focused down to ~10 μm in diameter using a 50× objective lens. The focused beam size is substantially smaller than an individual device (~20 μm × 20 μm), which allows us to selectively excite different topological states—among the bulk, edge, and corner states—by selecting an appropriate spot location for excitation. We excited nine different positions sequentially: one centre, four side edges, and four corners of the square-shaped topological boundary, which are depicted in the insets of Fig. 3(a). Figure 3(a) shows the microphotoluminescence (μPL) spectra measured, while the excitation spot was located at the nine different positions. All the spectra were obtained at excitation power levels above the lasing thresholds, resulting in sharp stimulated emission peaks from the corresponding topological eigenstates. The bulk exhibits a single-mode lasing action despite the absence of active optical confinement, a characteristic of band-edge laser[40,41]. Meanwhile, the emission spectra from the edges exhibit multiple modes, which we attribute to Fabry–Pérot oscillations along the side edge terminated by two corners. The effective refractive index for the oscillation modes can be estimated from the equation for the free spectral range, $\Delta\lambda = \lambda^2/(2n_{eff}L)$, where $L$ is the physical cavity length. Using the experimentally determined modal separations $\Delta\lambda \approx$ 2–3 nm and assuming that one side edge constitutes the Fabry–Pérot cavity $L$ = 10 μm, we obtained $n_{eff}$ = 36–59. Such a large effective refractive index indicates that edge state lasing occurs near the maximum of the edge state dispersion curve where the photon group velocity becomes smaller—see the solid blue line in Fig. 2(b). Regarding the corner excitations, all the spectra exhibit single-mode lasing actions, indicative of strong light confinement. Although the lasing wavelengths fluctuate, which is presumably due to structural inhomogeneity caused by imperfect nanofabrication, the average values are well distinguished and ordered sequentially for the three topological states: as we approach the higher-order TESs (from the bulk to the edge to the corner states), the lasing peaks shifted to blue. This experimental observation is qualitatively consistent with the relative energy positions of the corresponding topological states, as identified in the projected band structure—Fig. 2(b). The preservation of relative spectral positions of the three topological states is an indirect but strong evidence for their topological nature and modal robustness. Furthermore, it may imply that our bulk state lasing originates from the band-edge of the lower bulk band at $a/\lambda \approx 0.32$, which corresponds to the first M-point band edge, as identified from the band structure in Fig. 1(b).

**Simulations on topological states**

We conducted FDTD simulations to theoretically confirm the eigenstates supported by the structure. Several randomly oriented dipole sources were deployed at appropriate locations, while monitors were set above the slab to record signals emitted from the corresponding area. Simulation results are summarised in Fig. 3(b). It is noteworthy that the spectra from the four edges as well as those from the four corners are identical, which is due to the 4-fold rotational symmetry of the simulation model structure. The bulk and edge states support multiple modes whereas the corner state is singular. The multiple modal nature of the bulk state can be explained by the finiteness in sample size, which effectively lifts the degeneracy of an otherwise single band-edge mode in the ideal infinite PhC structure; the modal profiles that correspond to the three bulk modes in Fig. 3(b) are shown in the Supplementary Information. Among them, only the shortest wavelength mode is likely to lase because its modal shape matches best with the Gaussian optical excitation profile and because its spectral position overlaps the most with the optical gain bandwidth. Regarding the edge modes, the Fabry–Pérot modal separations, 2–3 nm, are qualitatively consistent with the experimental observation. The modal profiles for a few representative edge modes are calculated and displayed in the Supplementary Information. The overall spectral blue-shift as the topological order increases is consistent with the experimental results in Fig. 3(a).

To consider realistic situations such as random structural variations caused by imperfect nanoscale fabrication, we intentionally introduced disorder to the model structure by randomly varying the individual hole size to within ±10% of the design value. As shown in Fig. 3(c), the realistic model structure removes the four-fold degeneracy from the edge and corner states. The number of edge modes is significantly reduced, while the dominant mode becomes unpredictable, which qualitatively matches well with the experimental results. Regarding the corner states, the single mode nature is preserved well despite the introduction of structural disorders. Similarly, the global spectral blue-shift from the bulk to the edge to the corner states is well preserved, confirming the topological nature of the states.

**Emission patterns**

Figures 3(d)–(f) show the images of emission patterns obtained by both experiments and simulations. The experimental images in Fig. 3(d) were captured using an InGaAs infrared camera, whereas the images in Figs. 3(e) and 3(f) corresponded to Figs. 3(b) and 3(c), respectively, which were obtained from FDTD simulations. The 2D, 1D, and 0D natures of the corresponding topological states are clearly reflected in the emission patterns, which can be summarised as an extended/dispersed modal pattern over the entire inner square region in the 2D bulk state, line-shaped patterns along the side edges in the 1D edge states, and single dot patterns isolated at the four corners in the 0D corner states. Furthermore, additional dot-like emission patterns are shown in the experimental bulk and edge state emission patterns—Fig. 3(d), which are seemingly different from the simulated ones shown in Figs. 3(e) and 3(f). However, the dot-like emission patterns are not associated with the corresponding eigenstates but are due to the out-couplings of the bulk and edge state emissions at the corners, which occur in the M and X directions from the viewpoints of the bulk and edge states, respectively. The dot-like emission patterns far away from the excited regions in Fig. 3(d)—the emission from the four corners in the bulk excitation case and the emission from the upper right corner in the left edge excitation case—supports our speculation strongly.

**Laser characteristics**

To assess the application possibility of the hierarchical topological states, especially of the higher-order corner state, as quality optical resonators, we examined their lasing characteristics. Fig. 4(a) shows the light-in versus light-out (*L–L*) characteristics observed from the four corner states. They exhibit similar lasing thresholds, 7–8 kW/cm$^2$, which is comparable to those of L3 PhC cavity lasers that we have fabricated previously using the same MQW epistructure[20]. The *L–L* characteristics of the edge and bulk states were measured (see Supplementary Information), and their laser thresholds were similar. These observations are consistent with the simulation results: quality factors simulated for the realistic structures—with structural fluctuations included—are similar (~10$^3$) for all the topological states (see also Supplementary Information). Figure 4(b) shows the polarisation distribution patterns measured in the far-field geometry for the four corner states, in which all exhibit a strong linear polarisation dependence along the 45° directions from the *x*- or *y*-axis. This observation matches with the net polarisation at the corner, which results from the vector sum of two orthogonal edge polarisations, as illustrated in Fig. 2(d).

**Hierarchical corner states**

Although FDTD simulations and μPL measurements unambiguously proved the simultaneous presence of 0D and 1D TESs as well as 2D bulk states in our structure, the origin of their stability and robustness as optical modes still remains unclear. Hence, we performed analytic model calculations. SSH model structures are often oversimplified in tight-binding calculations such that the interaction strength is nonvanishing only between the nearest neighbours. The corresponding—thus unrealistic—situation for our tetramerised PhC structure is depicted in Fig. 5(a), where there are only two kinds of the nearest neighbours (or interaction strengths). Such oversimplified model calculations preserve the chiral symmetry and yields a symmetric eigenvalue spectrum with respect to zero energy[28,39]. Figure 5(b) summarises such oversimplified calculation results on our structure, in which the corner states are embedded within the symmetric bulk band. Although the tight-binding calculations proved the existence of the corner states, the resultant corner states are expected to be vulnerable because they can easily and inevitably hybridise with the degenerate bulk states. Figure 5(c) shows the accumulated spatial intensity distribution of the wavefunctions associated with the four corner states, which clearly indicates that the corner states are strongly coupled with the bulk states.

For comparison, we constructed a Hamiltonian matrix that is more realistic and appropriate for our PhC-based structure, where the hopping probability (or interaction strength) remains finite beyond the nearest neighbours; we assume in this model calculation that the hopping probability decays exponentially as a function of separation distance between interacting atoms—Fig. 5(d). Such long-ranged and complex interactions effectively break the chiral symmetry and modify the band structure such that the corner states move out of the band and fall inside the bandgap[42], as shown in Fig. 5(e). Owing to the lack of a degenerate bulk state to interact with, the resultant corner states exhibit the wavefunction distributions of stable modes—Fig. 5(f). These simple Hamiltonian model calculations suggest that stable in-gap higher-order TESs can be obtained by a chiral symmetry breaking even when bulk multipole polarisation does not exist. Repeated simulations with different interaction parameters confirmed the importance of the corner states inside the bandgap for modal stability. The robustness of the in-gap corner state in the presence of an intentional structural defect is discussed in the Supplementary Information.

**Discussion**

In summary, we designed and fabricated a topologically engineered 2D PhC platform using an InP-based semiconductor MQW epilayer to demonstrate lasing actions at all topological states available in the structure. Spatially resolved optical excitations resulted in lasing actions not only at the 2D bulk state in the central region, but also at the TESs at 1D edges and 0D corners. All these topological lasers exhibit sharp lasing peaks and clear laser thresholds. A direct imaging of emission patterns proves that lasing actions occurred at the corresponding topological states. These experimental demonstrations provide insights into the development of not only sophisticated but also robust topological photonic integration platforms. In addition, our topological platform may serve as a powerful testbed for various fundamental studies regarding higher-order topological photonic states, which may enable totally new topological photonics when combined with nonlinearity[43], non-Hermitian physics[44], or quantum photonics.

## Methods

### Numerical simulation

The band structure calculations and mode profile simulations were conducted using a commercial FDTD software (FDTD solutions, Lumerical). For all the calculations, a full three-dimensional simulation formalism was employed. The refractive index of the InGaAsP slab was assumed to be real and constant ($n = 3.4$), which was reasonable at high excitation levels where the gain materials became transparent and for the limited frequency range considered in this study. For band structure calculations, the Bloch boundary conditions were applied in the directions of periodicity. For the spectrum calculations shown in Fig. 3, monitors were deployed in the plane parallel to the sample surface to collect light emitting in the vertical direction. For the tight-binding calculations shown in Fig. 5, a $6 \times 6$ array of unit cells was considered. The hopping or interaction strengths ($h$) between photonic atoms was determined by a single exponentially decaying function, $h = \exp(-s)$, where $s$ is the interatomic distance. The long and short distances between the nearest neighbours were arbitrarily selected to be 3.5 and 1.5, respectively, while all on-site energies were set to zero.

### Device fabrication

An epi-wafer for device fabrication was procured commercially, in which an InP buffer, InGaAs etch-stop layer, 1-µm-thick InP etch-sacrificial layer, and 230-nm-thick InGaAsP MQW membrane were grown sequentially on top of an InP substrate. A 30-nm-thick $Si_3N_4$ hard mask layer was deposited using plasma-enhanced chemical vapour deposition (Plasma-Therm, Inc.). Subsequently, electron-beam resist (ZEP520A, Zeon Corp.) was spin-coated and patterned to define the PhC structures using e-beam lithography (Raith GmbH). After developing the patterned resist, $CH_4/H_2$-based reactive-ion etching (Oxford Instrument) was employed to transfer the patterns to the underlying MQW membrane. Finally, the InP etch-sacrificial layer was wet-etched in HCl to complete device fabrication.

**Data availability**. The data sets within the article and Supplementary Information in the current study are available from the authors upon request.

## Acknowledgements

This work was supported by Samsung Research Funding Center of Samsung Electronics under Project Number SRFC-MA1801-02

## Additional information

**Supplementary Information** accompanies this paper at

**Competing interests:** The authors declare no competing interests.

**Reprints and permission** information is available online at

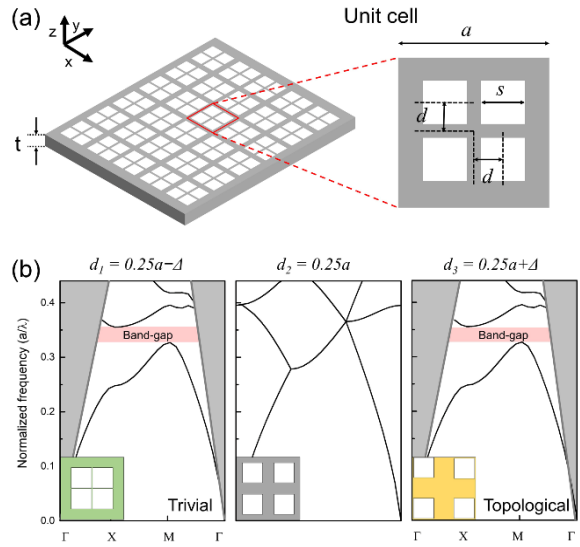

Fig. 1. Topological 2D PhC structure. (a) Schematics of the 2D square-lattice PhC slab composed of an array of air-hole tetramers. The unit cell of lattice constant $a$, enclosed by solid red lines, is composed of four square-shaped holes with side length $s$ and separation distance $2d$. (b) Calculated TE-polarised band structures of the PhC slab with different $d$ values: $d_1 = 0.25a − \Delta$, $d_2 = 0.25a$, $d_3 = 0.25a + \Delta$, where $\Delta = 0.07a$. The regions shaded in grey and pink represent light cones and photonic band-gaps, respectively.

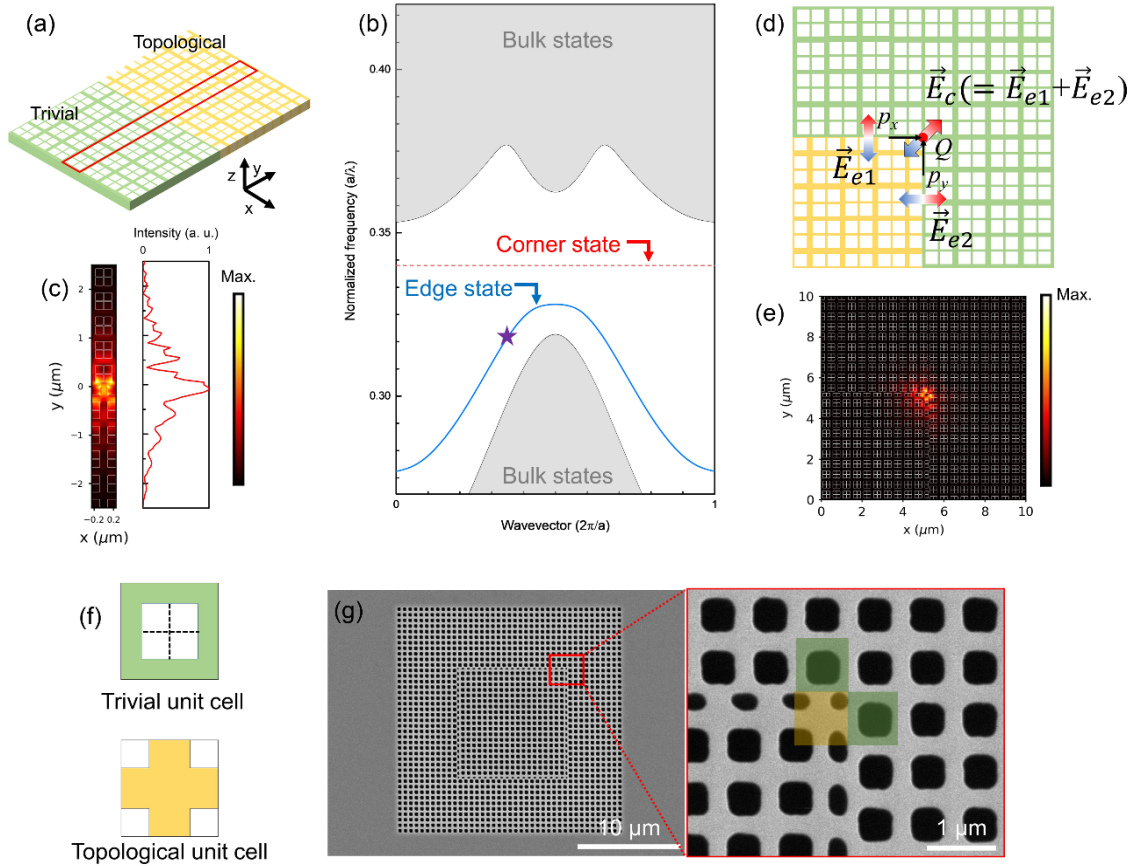

Fig. 2. Topological boundaries and associated eigenstates. (a) Schematics of a straight edge boundary between topologically nontrivial (yellow) and trivial (green) PhC structures. Enclosed by the solid red lines is the unit cell used for numerical simulations. (b) Projected band structure calculated for the PhC structures shown in (a). The regions shaded in light blue represent the bulk states, whereas the solid blue line inside the bulk band-gap is the dispersion relation of the 1D edge state. (c) Simulated modal profile of an edge state ($k = 0.35 \times 2\pi/a$), which is marked by a purple star. (d) Schematics of a corner boundary where the topologically nontrivial PhC is surrounded by the trivial one. An effective polarisation charge $Q$ is accumulated at the corner where two edge dipole polarisations, $p_x$ and $p_y$, combine. The frequency of the resultant corner state is indicated by the dotted red line in (b). (e) Simulated modal profile of the 0D corner state. (f) Unit cell designs for experimental realisation of the topologically trivial (top) and nontrivial (bottom) PhCs. The white areas are the air-holes. (g) SEM images of the fabricated device: the entire structure (left) and upper right corner boundary (right). The dark areas are the air-holes. The scale bars are 10 μm (left) and 1 μm (right).

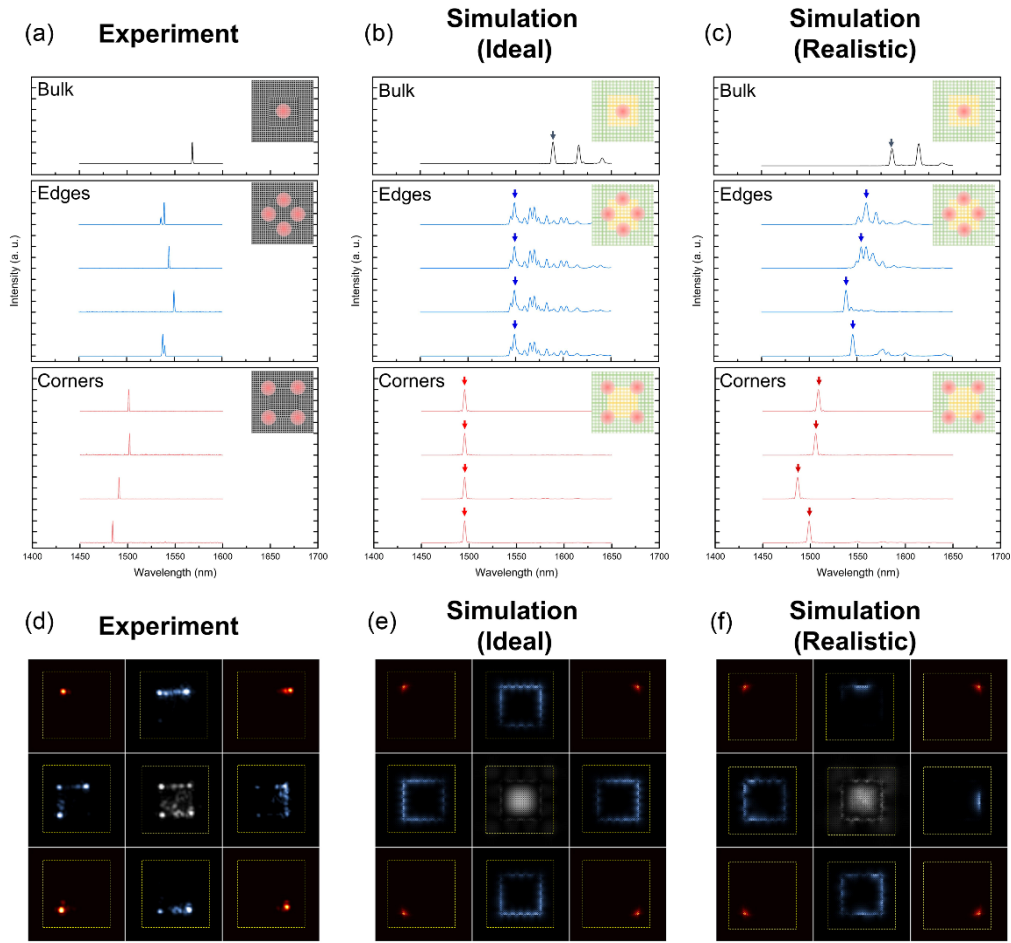

Fig. 3. Lasing from the 2D, 1D, and 0D topological eigenstates. (a) μPL spectra obtained when the optical excitation spot was at (from top to bottom) the bulk, edges, and corners. The insets schematically identify the positions of the excitation spot (red circles) on the device. (b) Modal spectra simulated for a perfect model structure with the four-fold rotational symmetry. Randomly oriented dipole clouds were deployed at the positions indicated by the red circles in the insets. (c) Modal spectra simulated for a realistic structure with disorder. (d) Laser emission patterns obtained from above the sample. Each image was captured using an infrared camera, while the corresponding position (one bulk, four edges, and four corners) was excited. (e)(f) Simulated modal profiles of the bulk, edge, and corner modes. When modes are multiples, as in the cases of the bulk and edge modes, the presented modes are selected among the dominant ones, which are marked by the arrows in (b)(c).

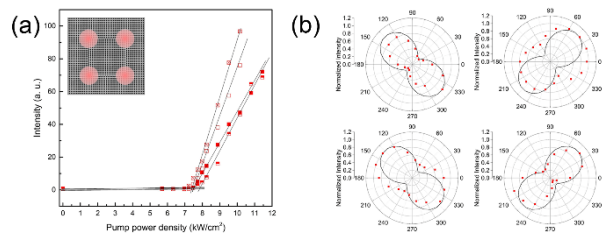

Fig. 4. Characteristics of the corner state lasers. (a) Light output intensity as a function of excitation power density, measured for the four corner states in a device. (b) Far-field polarisation dependences of the corner state lasers. The red dots are measured data, while the black lines are from FDTD simulations.

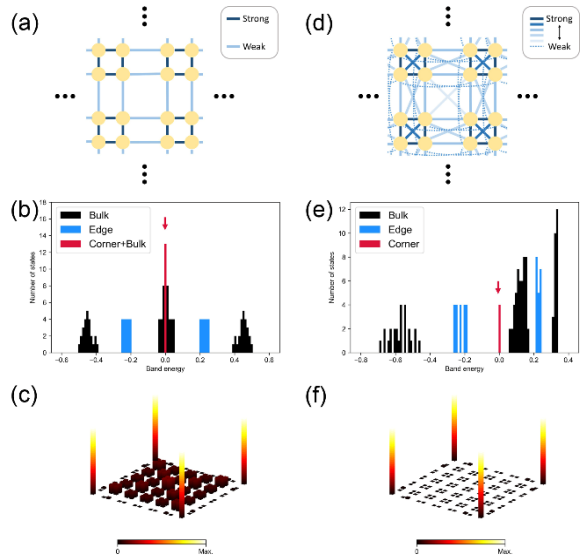

Fig. 5. Formation and stability of the 0D corner states. (a) Tetramerised 2D PhC structure with only two nearest-neighbour interactions considered. The yellow dots and the blue connecting lines between them represent photonic atoms and interaction strengths, respectively. Two different interaction strengths are expressed with different colour contrasts of the lines. (b) Calculated eigenvalue distributions for the model structure shown in (a). (c) Spatial distribution of the eigenstates at the energy marked by the arrow in (b). (d) Tetramerised 2D PhC structure with six different interaction strengths (or interatomic distances) under consideration. The six interaction strengths are distinguished by colour contrasts and shapes of the lines. (e) Calculated eigenvalue distributions for the model structure shown in (d). (f) Spatial distribution of the eigenstates at the energy marked by the arrow in (e).

## Supplementary Information

### Calculation of the polarization as a topological invariant

In the theory of polarization in solids [1], the bulk dipole moment $\boldsymbol{p} = (p_x, p_y)$ of a 2D crystal is given by

$$p_{x,y} = -\frac{1}{(2\pi)^2} \int_{BZ} d^2\boldsymbol{k} \, Tr[i\langle u_m(\boldsymbol{k})|\partial k_{x,y}|u_n(\boldsymbol{k})\rangle],$$

where *m* and *n* represent the indices of all the bands below the band-gap of interest. We numerically calculate this value of our system by using a finite-difference time-domain (FDTD) software. Note that numerical evaluation of derivatives and integrations with respect to discrete k points relates this calculation to Wilson-loop approach, which has been introduced in previous studies [2, 3]. It is also well-known that a mirror symmetry in $x(y)$ direction quantizes $p_x(p_y)$ to 0 or 0.5 [2, 3]. We first obtain eigenvectors $|u(k_x, k_y)\rangle$ by calculating the H$_z$ field distribution in the unit cell at each discrete $(k_x, k_y)$ point, and then numerically compute the bulk polarizations. Figure S1(a) shows the calculation results as a function of $d$. As expected, $p_x(p_y)$ values show a phase transition at $d = 0.25a$, where it jumps from 0 to 0.5, inferring that the 1D edge states can exist for $d > 0.25a$. Subsequently, the polarizations of the edge states can be calculated similarly to the bulk polarization; the results are plotted in Fig. S1(b). It turns out that the edge states polarization is also nontrivial, which in turn guarantees the existence of the corner states.

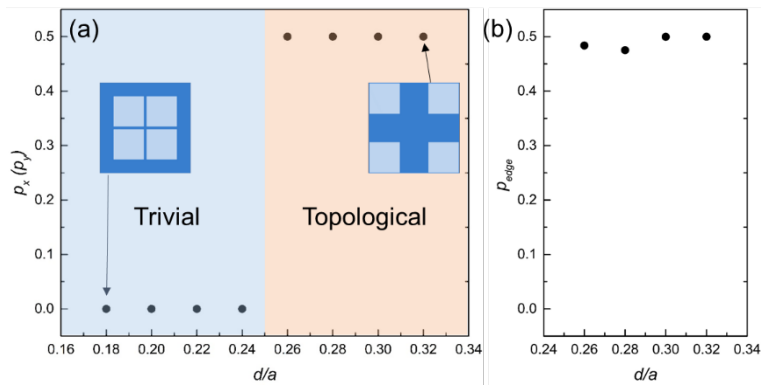

Fig. S1. Calculated polarizations as a function of *d/a* for (a) the bulk and (b) the edge.

**Performance characteristics of the topological state lasers**

To evaluate the qualities of the topological state lasers, light-in versus light-out characteristics were measured. Typical L-L curves for the nine different topological states, all measured from a single device, are shown in Fig. S2(a). Although the topological states are amply different, their lasing thresholds are rather similar, ~7 kW/cm$^2$. This experimental observation results are qualitatively consistent with the simulation results, which are shown in Fig. S2(b): the simulated cavity Q-factors are in the order of 10$^3$, regardless of the topological states. The Q-factors are calculated for the modes shown in Fig. 3(c) in the main manuscript, which belong to the realistic model structure with disorder.

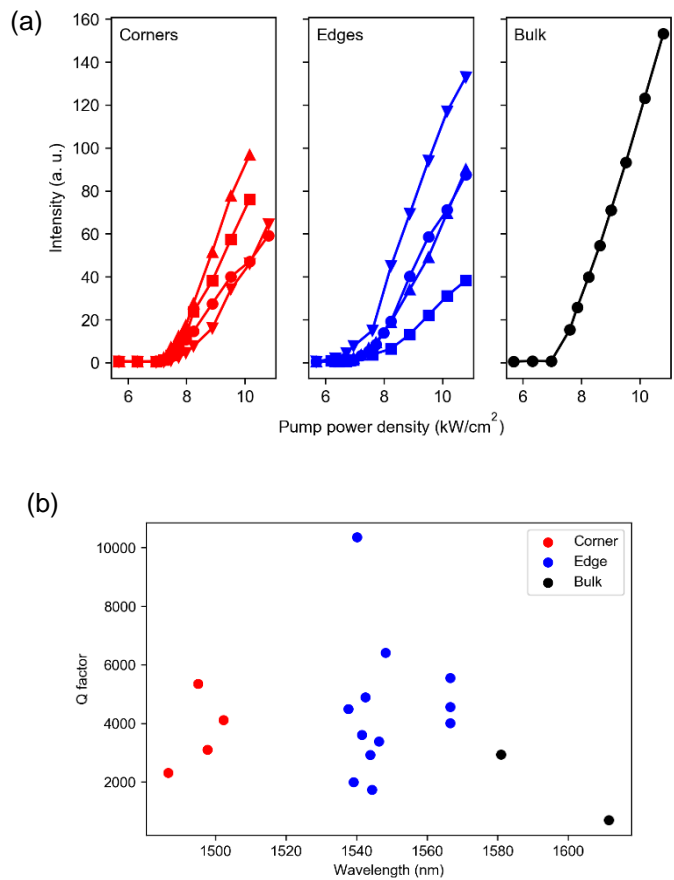

Fig. S2. (a) Relationships between light output intensity and excitation power density for the corner, edge, and bulk states. (b) Quality factors calculated for the simulated topological modes.

**Identification of the bulk states**

An ideal band-edge laser structure should be infinitely periodic. In reality, however, a band-edge laser is finite. When the excitation area is large enough, the device boundaries affect the formation of normal modes and lift the modal degeneracy of the ideal band-edge laser, as demonstrated by FDTD simulations—see Fig. 3(b) in the main manuscript, which is copied here as Fig. S3(a). In order to prove that those three eigenmodes are indeed originated from a band-edge mode, we calculate their modal profiles, which are shown in Fig. S3(b). The shortest-wavelength mode (Mode 1) resembles the ideal band-edge mode most, while the other two (Mode 2 and 3) possess the modal patterns of higher-order modes. For further assessment, we take Fourier-transformations of the modal profiles—Fig. S3(c). It is now inarguably clear that all the three modes are originated from the M-point band-edge mode.

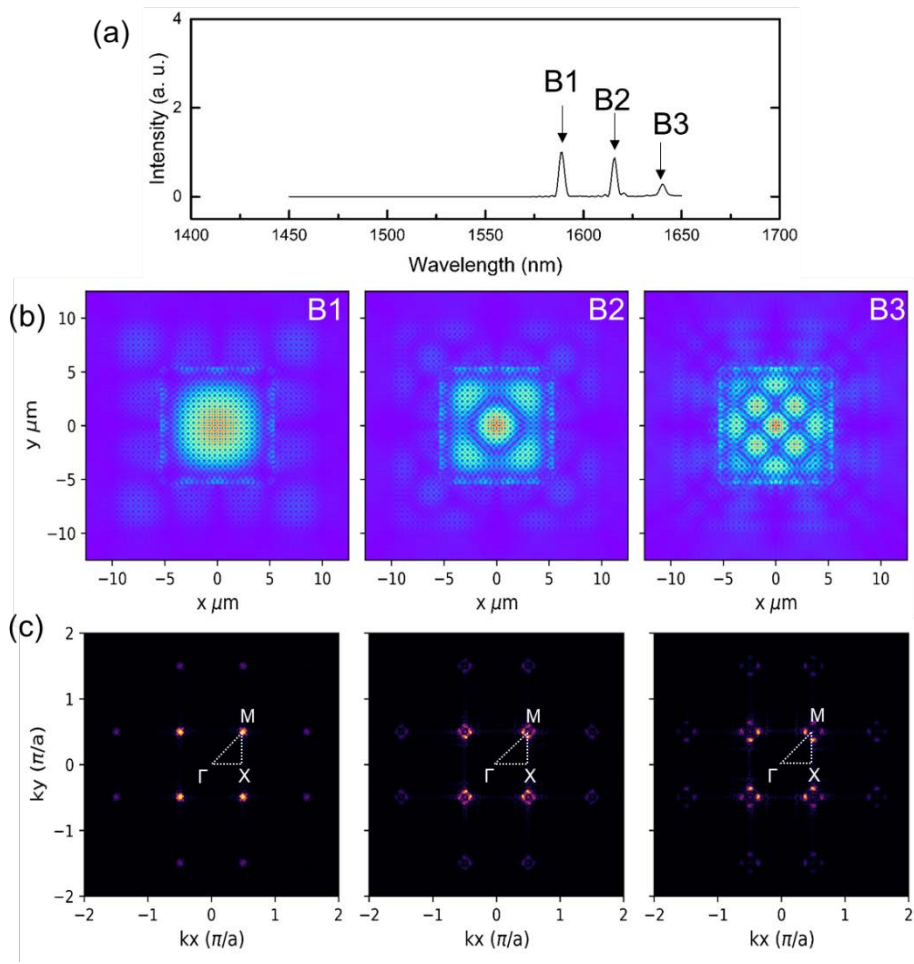

Fig. S3. (a) FDTD-simulated modal spectra of the bulk states. (b) Mode profiles of the three bulk states shown in (a). (c) Fourier-transformation images of the three bulk mode profiles in (b). The irreducible Brillouin zone is inserted for reference.

**Modal profiles of the edge states**

In order to clarify the physical characteristics of the edge state lasers, we perform FDTD simulations. Figure S4(a) is the replot of a part of Fig. 3(b) in the main manuscript, the simulated modal spectrum of the edge states in our topological PhC structure. Shown in Fig. S4(b) are the mode profiles of the three shortest-wavelength modes, which are identified by the arrows in Fig. S4(a). One can see that the modal periodicity in each figure is commensurate with the underlying PhC lattice, which is modulated by an envelope function with a much longer period. We believe that the envelope function is a result of mode beating between multiple cavity lengths of $L$, $2L$, $3L$, and $4L$, where $L$ is the side length of the square boundary.

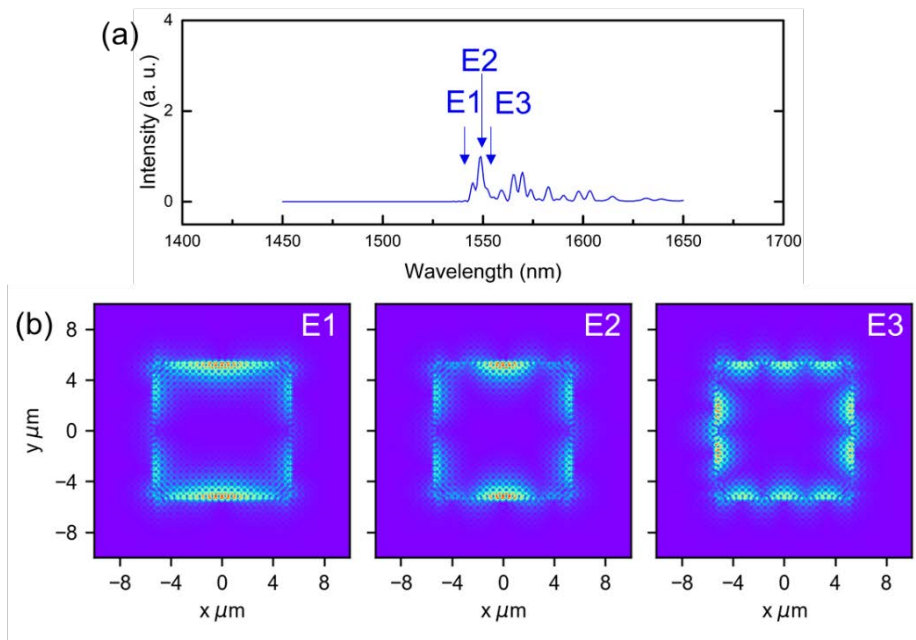

Fig. S4. (a) FDTD-simulated modal spectra of the edge states. (b) Simulated mode profiles of the three edge states with the shortest wavelengths, which are indicated by the arrows in (a).

**Modal robustness of the corner state**

Modal robustness of the corner state has been investigated in a few previous studies [4, 5]. In the present work, we examine an extreme situation where an edge defect, a missing atom from a tetramer, is very close to a corner. Despite the existence of such a strong defect near the corner, the corresponding corner state is still well defined with the modal profile rarely disturbed, as can be seen in Fig. S5. The calculated Q-factors are ~3,100 and ~2,500 without and with the defect, respectively.

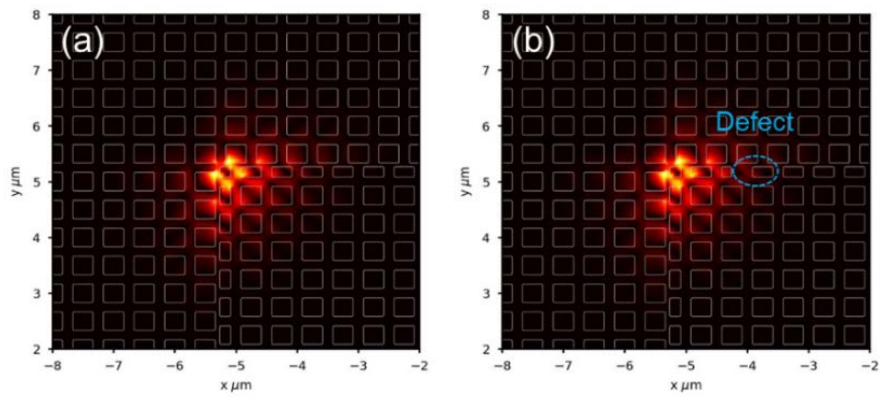

Fig. S5. Robustness of the corner state. (a)(b) Simulated electric field profiles of the corner states (a) without and (b) with a defect.